\documentclass[conference]{IEEEtran}
\IEEEoverridecommandlockouts
\usepackage{mathtools}
\usepackage{cite}
\usepackage{amsmath,amssymb,amsfonts}
\usepackage{algorithmic}
\usepackage{graphicx}
\usepackage{textcomp}
\usepackage{xcolor}
\usepackage{lipsum}
\usepackage{cuted}
\usepackage{array}
\def\BibTeX{{\rm B\kern-.05em{\sc i\kern-.025em b}\kern-.08em
    T\kern-.1667em\lower.7ex\hbox{E}\kern-.125emX}}

\makeatletter
\def\ps@IEEEtitlepagestyle{%
  \def\@oddfoot{\mycopyrightnotice}%
  \def\@evenfoot{}%
}
\def\mycopyrightnotice{%
  {\footnotesize  978-1-6654-9071-9/23/\$31.00~\copyright2023 IEEE \hfill}% <--- Change here
  \gdef\mycopyrightnotice{}% just in case
}

\begin{document}

\title{Angular Stability Analysis of Parallel Connected Grid-following PV Inverters}

%{\footnotesize \textsuperscript}

\author{Ramkrishna Mishan,~\emph{Student Member, IEEE} \\ Department of Electrical \& Biomedical Engineering, University of Nevada, Reno, Reno, NV 89557 
\\(email: \ mishan@nevada.unr.edu)\vspace{-0.2ex}}

% \IEEEoverridecommandlockouts
% \IEEEpubid{\makebox[\columnwidth]{978-1-5386-5541-2/18/\$31.00~\copyright2018 IEEE \hfill} \hspace{\columnsep}\makebox[\columnwidth]{ }}

\maketitle

\IEEEpubidadjcol

\begin{abstract}

High penetration of distributed generators (DG) in modern power grids creates angle,  voltage, and frequency instabilities.  
Most of the work in the literature has focused on small-signal stability analysis of single grid-connected inverters without thoroughly investigating their transient stability for large disturbances and interactions between parallel-connected inverters. To address these challenges, this paper examines the transient angular stability of a cluster of grid-following current source inverters. In a low inertia weak grid environment, grid-following inverters may lose synchronism due to faults and large-signal disturbances. A comparison of the voltage injection angle for pre-fault, fault-on trajectory, and post-fault conditions is performed, and an evaluation of the critical fault clearing time is directly related to the uniformity of the aggregated grid-following inverters. The results show that some inverters lose synchronism more quickly than others due to considerable variances in line impedances and apparent power values. Additionally, inverters with high apparent power are more susceptible to losing synchronism due to the maximum current limit, causing the voltage at the point of common coupling to decrease.
 
\end{abstract}


\section{Introduction}
The power grid structure and behavior are transforming rapidly as the percentile of low/no inertia solar photovoltaic (PV) inverter integration rise due to the depletion of conventional energy sources and recent climate change. Therefore, modern power systems are transitioning into weak grids with low inertia and less unit commitment\cite{9939706}, which has sparked recent research on islanded and grid-connected operation of PV inverters. Inverters of distributed energy resources (DERs) can be categorized into three types depending on the control strategy: Grid-following, Grid-forming, and Grid-supporting \cite{9184033}. Parallel-connected grid-following inverters are one of the most common types. However, these inverters are prone to losing synchronism in large-signal disturbances due to lack of inertia. Maximum power point tracking (MPPT) has been the typical control strategy for these inverters, and the strategy uses a phase-locked loop (PLL) to track the grid voltage of a common distributed node \cite{wang2020transient}. This node voltage is known as the point of common coupling voltage ($V_{PCC}$).   

In the angular stability assessment of traditional power grids, generators' voltage dynamics tend to be slower than the angle and frequency dynamics \cite{podmore1978identification}. As a result, generator voltages have been assumed constant for angular stability analysis. However, in weak-inertia grids, voltages of distributed generators change rapidly \cite{hosseinzadeh2021voltage}. Hence to maintain the stability of the power system, controlling the $V_{PCC}$ is widely used for the transient case of small-signal disturbances \cite{8975325}. Angular stability of inverters depends on the $V_{PCC}$ because the distribution system has a lower $X\slash R$ ratio than the transmission system. The traditional swing equation adopted for transient stability uses only active power because transmission systems have high $X\slash R$ ratio. Accordingly, the reactive power and voltage dynamics are inherently discarded. However, since distribution systems in a weak-gird condition have lower reactive power support, apparent power and voltage dynamics should be considered in the stability analysis \cite{8905556}. For parallel connected grid-following current source inverters, the magnitude and angle of the $V_{PCC}$ determine the current injection and power angle of the inverter.

Several methods have been developed to analyze the stability of DER inverters. In \cite{iyer2010generalized}, the authors have assumed the uniformity of multi-inverters to make final equivalent inverter differential-algebraic equations for the droop control laws. A Lyapunov energy function has been used in \cite{wang2017large} to analyze the rotor angle stability of interconnected multi-inverter power grids. In that paper, the authors assumed the virtual inertia of the grid-forming voltage source inverter. In contrast, household roof-top inverters are generally grid-following current source inverters. The authors of \cite{taul2020reduced} employed an aggregated reduced-order inverter model for wind farms with different transformer leakage impedances for synchronization by assuming the same dynamics and configuration of each inverter. The authors in \cite{raman2019mitigating} illustrated how the line parameters impact the multi-inverter stability after considering mutual interactions and coupling effects in a stiff grid condition. These papers overlooked the non-uniformity and the dynamics of inverter voltage during fault conditions. However, generalizing the inverter models may not provide accurate stability conditions for aggregated multi-inverters. After a certain period, the inverter ratings and line parameters change along with the weather changes. This paper utilizes the non-uniformity of parallel-connected inverters in terms of line impedances, virtual impedances, and apparent power to achieve a precise stability condition for PV multi-inverter systems. In a parallel-connected multi-inverter system, each inverter, together with synchronous generators, contributes to regulating the $V_{PCC}$. Furthermore, inverters' current injection depends on this common point voltage \cite{hassan2020review}. The peak current injection of solar photovoltaic (PV)-based inverter is intentionally limited due to the constraints in the power electronic components.

During the occurrence of faults in a weak grid, the Critical Clearing Times (CCTs) can be an indicator of relative stability. The CCT is defined as the maximum period between initiating and isolating a fault such that the power system remains stable. 

This paper develops a model for aggregated inverters' $V_{PCC}$. This paper form the individual inverter generation voltage, which depends on this $V_{PCC}$  and inverter parameters. Finally, we show the causality of reduced CCT of non-uniform inverter because of the maximum current limit. Inverters with large apparent power and a small X/R ratio are prone to lose synchronism faster than others. In such a case, the $V_{PCC}$ reduces further. Gradually, all connected inverters would lose synchronism if the fault is not cleared at that time.

The rest of this paper is structured as follows: Section II presents the mathematical modeling of an aggregated inverter model with virtual synchronization into the voltage loop. Section III discusses the reactive current limit causing angle instability. The results of case studies are addressed in section IV. Section V provides concluding remarks.

\section{Aggregated Inverter Model}
This paper employs the equivalent Thevenin network and superposition approach for multi-converter systems similar to what has been proposed in \cite{taul2020reduced}. The authors of \cite{taul2020reduced} have investigated a wind turbine multi-converter system with uniform converter parameters. However, we have considered a PV multi-inverter system with non-uniform inverter parameters in this work. In this type of system with multiple non-uniform inverters, the loss of synchronism occurs more rapidly, so to maintain stability of such a network, it is necessary to investigate the dynamics of aggregated inverter network. First, we determine the grid Thevenin equivalent voltage and the Thevenin equivalent impedance seen by the aggregated inverter network. This equivalent network data could be read by modern PMU \cite{7284716}. Then, we employed a virtual impedance in the voltage loop to synchronize correctly with the point of common coupling ($PCC$) for different inverter line impedances\cite{9813798}. With virtual synchronization process, we minimize the circulation current and improve reactive power-sharing among parallel-connected inverters. Both self-synchronization and controlled- synchronization implement in the voltage loop of the inverter.

\subsection{Thevenin Equivalent Network}
\begin{figure}[tbp]
\centering
\includegraphics[width=8cm]{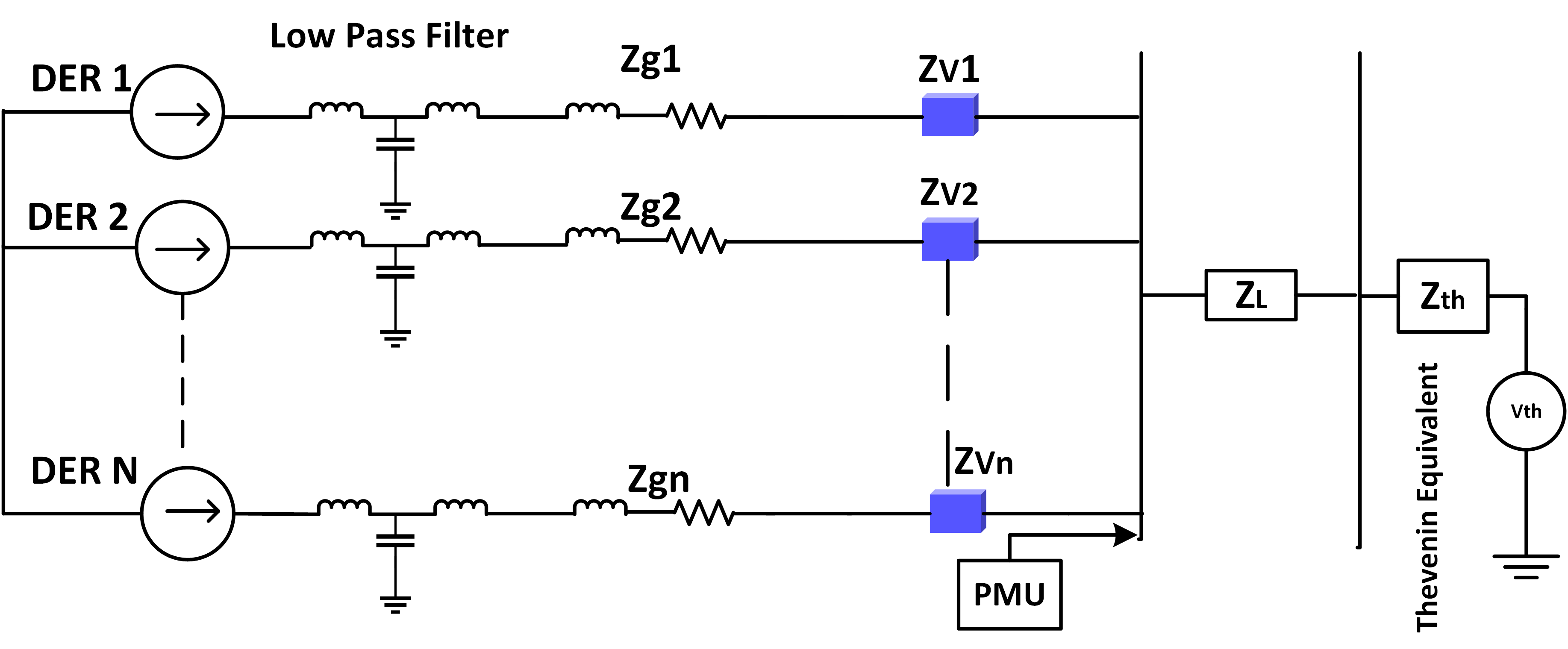}
\caption{Aggregated Current Source Multi-inverters Connection with Parallel Thevenin Equivalent Feeder}
\label{fig:PLL_GFL}
\end{figure}

The Thevenin equivalent network (TEN) can reproduce the precise behavior of a certain network without requiring comprehensive modeling of that network. TEN methodology is an equivalent voltage and impedance representation of the rest of the network buses as seen by a particular bus. An abstract representation of the TEN in a reduced network has been practiced in transient rotor angle stability  \cite{8331866}. Assuming the parallel inverter dynamics are faster than the other grid network dynamics, the grid can be replaced by its TEN representing a single voltage source (Vth) in series with an impedance (Zth) as shown in Fig. \ref{fig:PLL_GFL}. TEN parameters are determined for pre-fault and during fault to estimate the voltage angle for parallel-connected inverters.

In Fig. \ref{fig:PLL_GFL}, the $V_{th}$ and $Z_{th}$ are determined before evaluating the dynamics of $V_{PCC}$ using the superposition theorem on the parallel inverters and the connected feeder. $Z_{gp}$ indicates the lines impedance and $Z_{Vp}$  denotes the virtual impedance for $P^{th}$ inverter.

\subsection{Aggregated Inverters Inner Voltage Loop Impacted by $V_{th} $ and  $V_{PCC}$}
In this study, the dynamics of the parallel inverter branch are more significant than that of the connected feeder branch; therefore, the $V_{th}$ is assumed constant during the stability analysis. However, implementing the superposition theorem will only be viable if utility lines and customer load impedances are known. In  \cite{taul2020reduced}, the $V_{PCC}$ is defined using the superposition of the linear network. In this paper, we modified the proposed equation considering how the magnitude of previous $V_{PCC}$ causes different current injections.
\begin{equation}\label{equ:inv_PCC}
\begin{aligned}
    v_{PCC}(t)=v_{th}+Z_{eq1} \times  \displaystyle\lvert \frac{S_1}{v_{PCC}(t-T_d)} \rvert 
        (\cos{\theta}_{cg1}+ j\sin{\theta}_{cg1}) \\+ \cdots +Z_{eqn} \times   \displaystyle\lvert \frac{S_n}{v_{PCC}(t-T_d)} \rvert (\cos{\theta}_{cgn}+j\sin{\theta}_{cgn})
\end{aligned}
\end{equation}
where $T_d$ is sampling time; $S_1$, $S_2$, $\cdots$,  $S_n$ are the injected apparent power of the inverter 1, 2, $\cdots$, $n$;  $Z_{eqp}=(Z_{gp}+Z_{vp})||(Z_{th}+Z_{L})$, $P$=1,2, $\cdots$, $n$
and power injection angel ${\theta}_{cgp}$; and  is the sum of angle extracted across $V_{PCC}$ voltage and power factor angle of inverter, $P$.

Here, $v_{PCC}(t)$ and $v_{PCC}(t-T_d)$  represent $V_{PCC}$ at time $t$ and ($t-T_d$) respectively. However, we adopt these two as identical in this paper by assuming that the PLL' ADC provides a very small sampling time. 

Putting (\ref{equ:inv_PCC}) in a compact form yields the following expression,
\begin{equation}\label{equ:inv_PCCs}
\begin{aligned}
    v_{PCC}=v_{th}+\sum_{i=1}^{n} \left( (Z_{eqi}) \times  \displaystyle\lvert \frac{S_i}{v_{PCC}} \rvert 
        (\cos{\theta}_{cgi}+j\sin{\theta}_{cgi}) \right)
\end{aligned}
\end{equation}

After specifying $V_{PCC}$, the next step is determining the individual inverter generation voltage. 

\begin{equation}\label{equ: inverter_POS}
    \begin{cases}
        v_{g1}=v_{PCC}+ \displaystyle\left\lvert \frac{S_1}{v_{PCC}} \right\rvert (Z_{g1}+Z_{v1}) \times 
        (\cos{\theta}_{cg1}+j\sin{\theta}_{cg1})
          \\
          \\
         v_{g2}=v_{PCC}+ \displaystyle\left\lvert \frac{S_2}{v_{PCC}} \right\rvert (Z_{g2}+Z_{v2}) \times 
        (\cos{\theta}_{cg2}+j\sin{\theta}_{cg2}) \\
        \vdots\\
           v_{gn}=v_{PCC}+ \displaystyle\left\lvert \frac{S_n}{v_{PCC}} \right\rvert (Z_{gn}+Z_{vn}) \times 
        (\cos{\theta}_{cgn}+j\sin{\theta}_{cgn}) \\
    \end{cases}       
\end{equation}

Rewriting (\ref{equ: inverter_POS}) for a general case considering the $V_{PCC}$ from (\ref{equ:inv_PCCs}), we get the following expression, 
\begin{equation}\label{equ: inverter_POSss}
\begin{aligned}
    v_{gp}=v_{th}+\sum_{i=1}^{n} \left(  (Z_{eqi}) \times  \displaystyle\lvert \frac{S_i}{v_{PCC}} \rvert 
        (\cos{\theta}_{cgi}+j\sin{\theta}_{cgi}) \right) \\
        + \displaystyle\left\lvert \frac{S_p}{v_{PCC}} \right\rvert (Z_{gp}+Z_{vp}) \times 
        (\cos{\theta}_{cgp}+j\sin{\theta}_{cgp})
\end{aligned}
\end{equation}

These individual and common coupling voltages of parallel inverters are utilized to determine the reactive current limit and angular trajectory.

The $q$ component of  direct-quadrature-zero transformation of $V_{PCC}$ can be expressed as follows, 
\begin{equation}\label{equ:inv_PCCs_f}
\begin{aligned}
    V_{PCCq}=V_{thq}+\sum_{i=1}^{n} \left(  \displaystyle\left\lvert {Z_{eqi}} \right\rvert  \times   \displaystyle\left\lvert \frac{S_i}{v_{PCC}} \right\rvert
        \times \sin({\theta}_{cgi}+{\gamma}_{i} ) \right) 
\end{aligned}
\end{equation}
where ${\gamma}_{i}$ component comes from the equivalent impedance (${Z_{eqi}}$) angle.

In the $dq$ frame, the quadratic component of individual inverter generation voltage is as follows,
\begin{equation}\label{equ: inverter_POSs}
\begin{aligned}
    V_{gpq}=V_{thq}+\sum_{i=1}^{n} \left(  \displaystyle\left\lvert {Z_{eqi}} \right\rvert  \times   \displaystyle\left\lvert \frac{S_i}{v_{PCC}} \right\rvert
        \times \sin({\theta}_{cgi}+{\gamma}_{i}) \right) \\
        + \displaystyle\left\lvert \frac{S_p}{v_{PCC}} \right\rvert \times \displaystyle\left\lvert Z_{gp}+Z_{vp} \right\rvert  \times \sin({\theta}_{cgp}+{\psi}_{p})
\end{aligned}
\end{equation}
assuming the apparent power,  line impedance, and virtual impedance of individual inverter remain the same during the fault-on-trajectory conditions. In this large-signal multi-inverter models, the virtual impedance is the synchronization term used in the voltage loop. Owing to page limit, we could not show the figure; however, the designed parameters are presented in Table-I.  

A symmetrical three-phase fault is created for a parallel-connected second feeder to investigate the angular stability of parallel connected grid-following inverters. Also, considering the slower dynamics of the first parallel feeder (higher inertia compared to multi-inverter network), $v_{th}^f$ and $Z_{th}^f$ remain the same during the faulted condition.

\section{Causality of Grid-following Multi-inverter Angular Stability}
Due to the inherent control of non-uniform grid-following current source inverters, the voltage of the common coupling decreases faster in the parallel feeder fault condition. Total short current injections in the point of common coupling by parallel inverters will be as follows,
\begin{equation}\label{equ:tot_cur}
\begin{aligned}
    I_{sc}(max) \le \sum_{i=1}^{n} I_{sci}=\frac{ \displaystyle\lvert S_1\rvert+\displaystyle\lvert S_2\rvert+ \cdots+\displaystyle\lvert S_n\rvert} 
        {V_{PCC}}
\end{aligned}
\end{equation}

\begin{table*}[h!]
    \centering
    \caption{ Inverter Parameters Design}
    \begin{tabular}{ | c | c| c| c |c | c | c| c | } 
\hline
Inverter Name & $S_{rated}$ & Line Resistance $(\Omega)$ & Line Reactance ($\mu$H) & X/R Ratio &  Virtual Resistance $(\Omega)$ & Inner $K_p$(V) & Inner $K_i$ (V)  \\ 
\hline
$Inv\ 1$ & $6$KVA & 0.15 &  40 & 0.1005 & 0.16 & $4.31\times 10^{-3}$ & 260  \\ 
\hline
$Inv\ 2$ & 9KVA & 0.30 & 45 & 0.0565 & 0.12 & $4.45\times 10^{-3}$ & 259 \\ 
\hline
$Inv\ 3$ & 8KVA & 0.25 & 50 & 0.0754& 0.06& $4.67\times 10^{-3}$ & 255  \\ 
\hline
$Inv\ 4$ & 12KVA & 0.35 & 60 & 0.0646 & 0.00& $4.76\times 10^{-3}$ & 265 \\ 
\hline
$Inv\ 5$ & 10KVA & 0.30 & 65 & 0.0817& 0.04& $4.57\times 10^{-3}$ & 255\\ 
\hline
\end{tabular}
    
    \label{tab:my_label}
\end{table*}

During the fault, the Thevenin equivalent voltage becomes lower compared to pre-fault voltage. The difference between pre-fault and during-fault Thevenin equivalent voltage becomes higher for a low inertia weak grid. In such condition, then individual inverter generation voltages in (\ref{equ: inverter_POSs}) decreases faster. From  (\ref{equ:inv_PCCs_f}), it is evident that if fault-on Thevenin equivalent voltage reduces then the $V_{PCCq}^f$ decrease as well. During pre-fault conditions, the different current injection will not affect the stability. But in faulted low-voltage $V_{thq}^f$ operation, some inverters might lose the synchronism faster than others. In that case, the $V_{PCCq}^f$ is further reduced, which is illustrated in section IV. After the feeder fault incidence, the Thevenin voltage of the parallel feeder is reduced, $V_{PCC}^f$ reduced from the pre-fault condition and inverter 1 tripped. Then the injected apparent power $S_1$ becomes zero. As shown in  (\ref{equ:inv_PCCs_f}), the $V_{PCCq}^f$ and $V_{thq}^f$ reduce further. In that case, the current injection will increase for other connected inverters. In that way, all of the inverters lose synchronism. If parallel-connected multi-inverters have large discrepancies in line impedance and apparent power, the CCT would be largely reduced. 

\section{Case Studies}
The configurations of all the inverters are presented in Table \ref{tab:my_label}, along with their apparent power, non-uniform line impedance, proportional and integral gain factors, i.e., $Kp$ and $Ki$. To achieve cross-synchronization, nominal virtual resistance was used. We set the fourth inverter size as 12KVA and the fifth inverter size to 10KVA as seen in Table \ref{tab:my_label} of parallel-connected inverter configurations. 

%In Fig. 2, it is shown that the fourth inverter loses angular stability first due to its maximum reactive current limit, which is discussed in the previous section. Angular instability occurs in power electronics-based inverters in a low inertia grid. Due to low inertia, the voltage support from the primary grid is absent here. 

\subsection{Uniform Vs Non-uniform Parallel Connected GFL Inverters Fault-on-trajectory}
\begin{figure}[h!]
\centering
\includegraphics[scale=0.205]{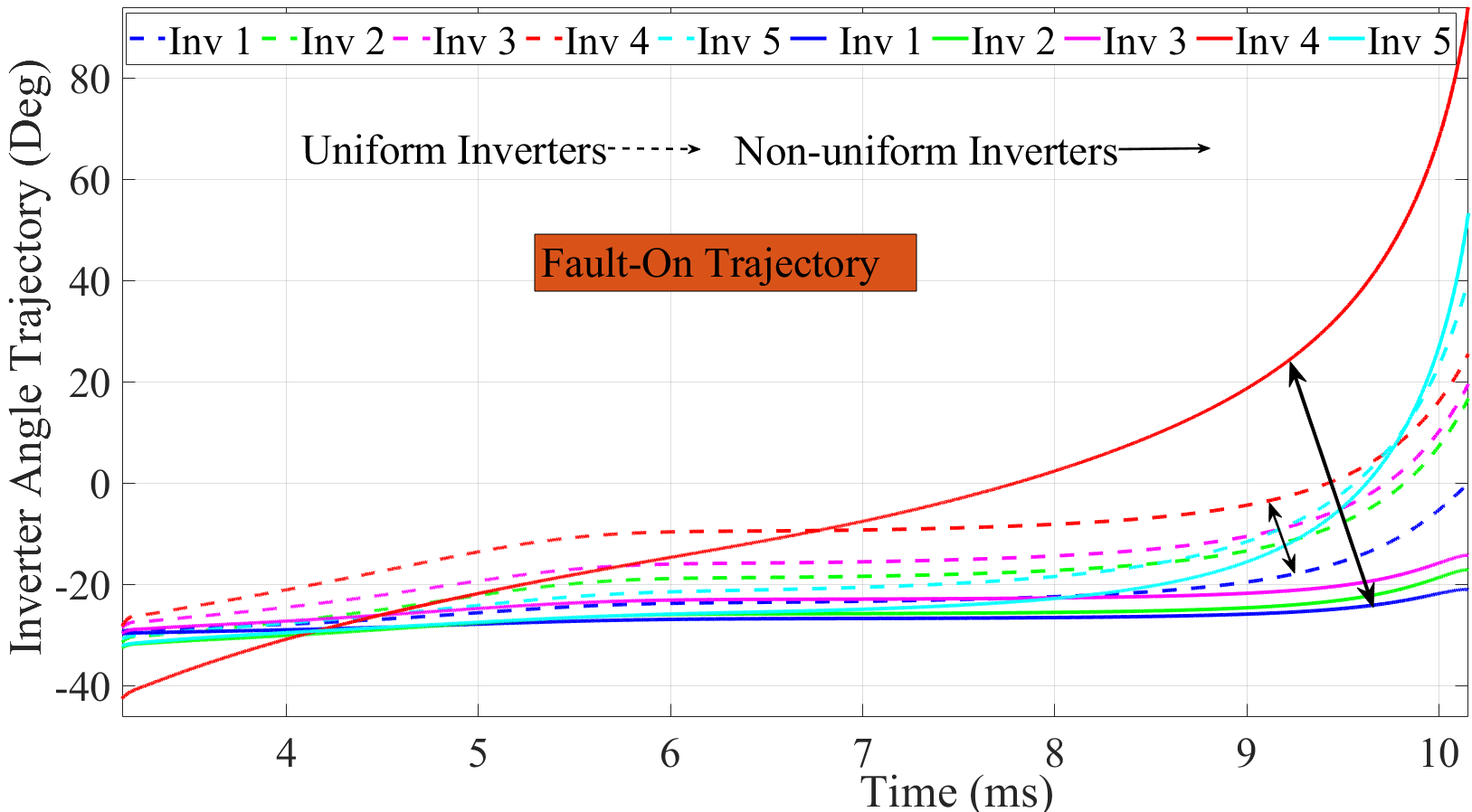}
\caption{Comparing nonuniform inverter with uniform inverter}
\label{fig:unu}
\end{figure}
Fig. \ref{fig:unu} illustrates the power injection angles trajectory of the parallel-connected non-uniform  verses uniform GFL inverters during the fault condition without clearing the fault. The Inverter 4, which has the highest power rating power lost synchronism faster than the others.
The inverter parameters that contribute to angular stability during large-signal disturbances is examined in this work. Fig. \ref{fig:unu} also shows the fault-on angular trajectory of uniform vs non-uniform parallel inverters. It shows that the angle trajectory of parallel-connected uniform inverters is influenced by the line impedances and the apparent power ratings. The dashed curve represents the angle trajectory of uniform inverters, while the solid line represent the angle trajectory of non-uniform inverters, with varying apparent power ratings. In comparison to non-uniform inverters, uniform inverters with identical power ratings exhibit angular instability in a small range. The current injection might increase because of the higher apparent power rating of one of the parallel-connected inverters. Still, the $V_{PCC}$ remains the same of these inverters. In that case, the inverter with high apparent power rating reaches the current limit faster than other inverters. As a result, it loses synchronism more quickly than the other which can be seen in Fig. \ref{fig:unu}. According to (\ref{equ:inv_PCCs_f}), the $V_{PCC}$ will reduce further. If we assume the same X/R ratio for all the inverters, the inverter will lose the synchronism in a sequence based on their apparent power rating. %second highest-rated inverter loses synchronism later.
 In this case, the critical fault clearing (CCT) time was significantly reduced for the non-uniform parallel-connected inverters.

\subsection{Non-uniform Inverters Fault-on-trajectory and Current Injection Limit}
\begin{figure}[h!]
\centering
\includegraphics[scale=0.205]{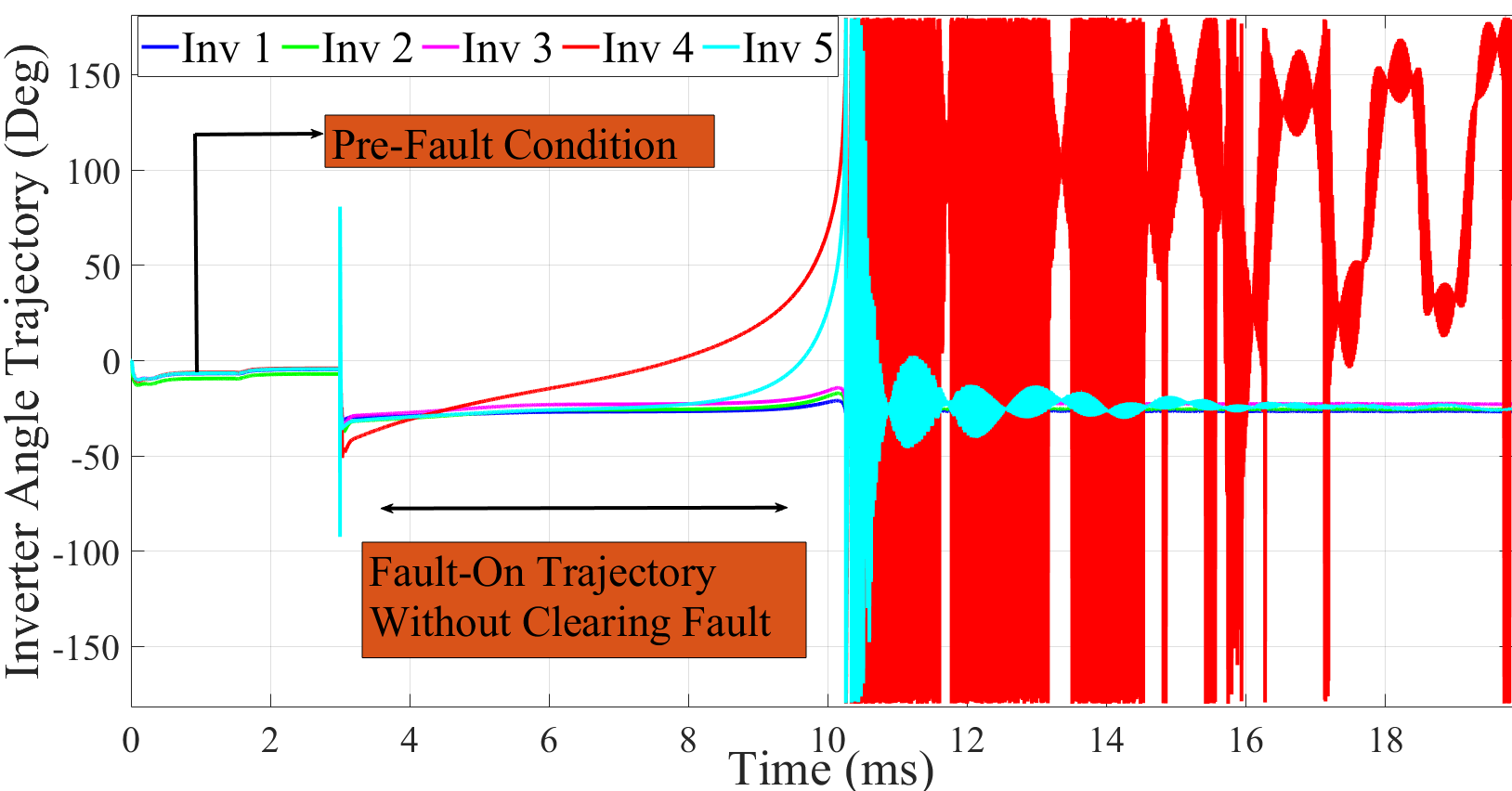}
\caption{Power Angle Trajectory of Parallel Inverters During Pre-fault, Fault-on Trajectory: Without Clearing the Fault}
\label{fig:awcf}
\end{figure}
\begin{figure}[h!]
\centering
\includegraphics[scale=0.205]{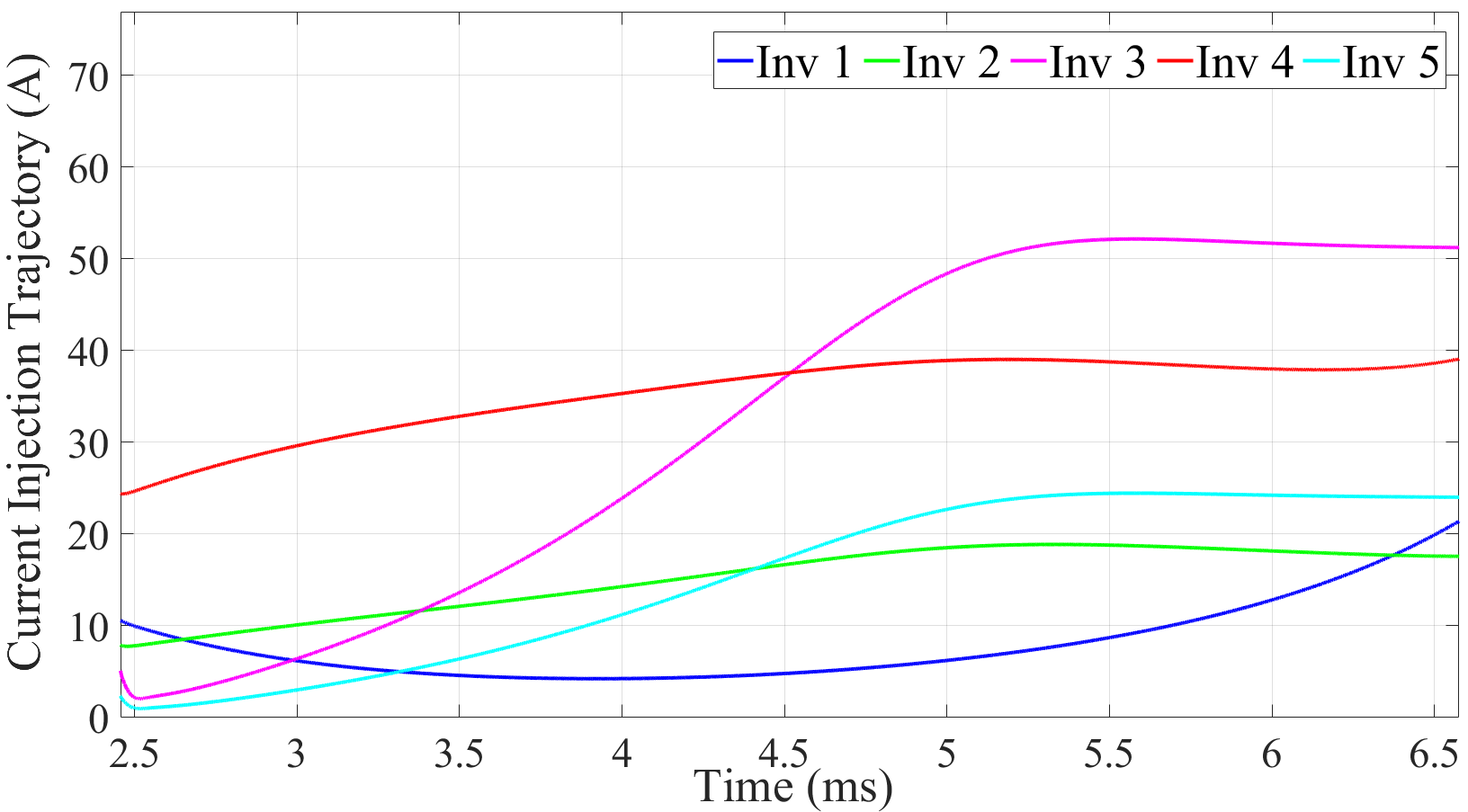}
\caption{Current injections for non-uniform inverter}
\label{fig:RCI}
\end{figure}

Fig. \ref{fig:awcf} shows the voltage angle trajectory of the parallel uniform inverters after creating a symmetrical line fault on a neighboring feeder without clearing it. All inverters are seen to lose synchronism after $10$ ms. 
Fig. \ref{fig:RCI} shows five inverters increasing reactive current before reaching the threshold value. In that case, the current limit is reached, and the inverter starts to lose synchronism as the grid needs more reactive power support, regardless of the active power and frequency conditions of the grid.

\subsection{Stable Case for Non-uniform Parallel Connected GFL Inverters, Clearing Before CCT}
In the following case, we develop a stable post-fault angle trajectory for these inverters. The associated feeder fault triggers around $t=3$ ms and clears after $t=4$ ms. We discovered that with this $1$ ms clearing time, post-fault angle revert to their original pre-fault angles after $t=4.8$ ms 

\begin{figure}[h!]
\centering
\includegraphics[scale=0.205]{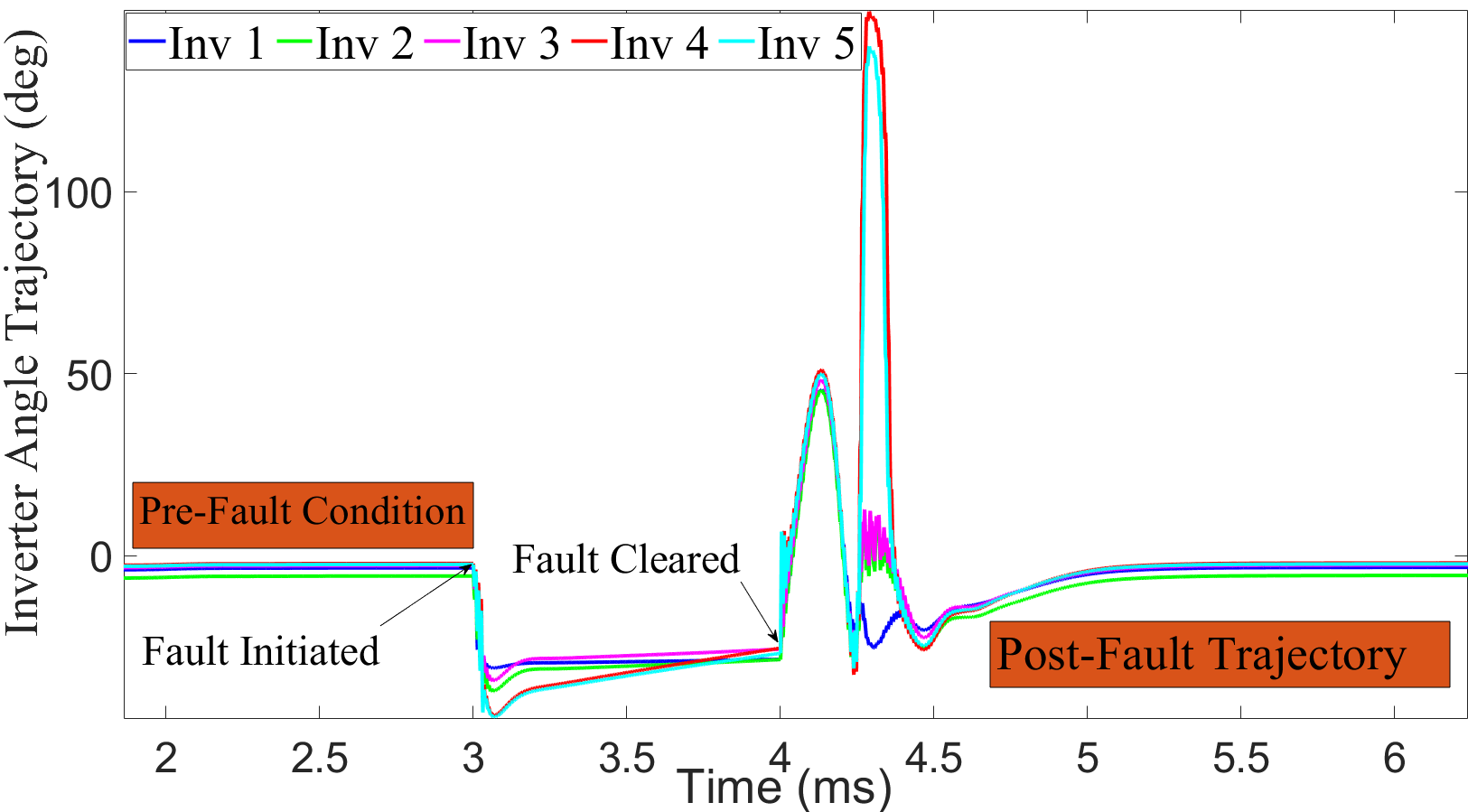}
\caption{Power Angle Trajectory of Parallel Inverters During Pre-fault, Fault-on and Post-fault conditions: Stable Condition}
\label{fig:at}
\end{figure}

Fig. \ref{fig:at} shows the inverter's voltage angle trajectory after clearing the fault 1ms after it occurred. At the instant of the fault clearance, the power angle swings before settling to a stable point where the pre-fault and post-fault voltage angles are same. The voltage angle deviation that happens during the line fault, as shown in Fig. \ref{fig:at}, restricts the peak current of the inverters. To comply with the reactive power injection, the reactive current increases during this $V_{PCC}$ drop. Due to the maximum current threshold in the dc side inductor, the peak current limit of the inverter is unavoidable. The reactive current keeps increasing during the fault-on-trajectory, as seen in Fig. \ref{fig:RCI}, until it reaches a maximum by the dc side inductors. The reactive current injection slope during fault-on-trajectory is directly related to the $X/R$ ratio of the inverter. For the same apparent power rating, an inverter with a higher $X/R$ percentage will have a higher slope during this increase. 

\subsection{Unstable Case for Non-uniform Parallel Connected GFL Inverters, Clearing After CCT}
In this case, we cleared the fault at $t=5$ ms, and discovered that the post-fault angles never return to pre-fault conditions, and it becomes completely unstable after $t=5.21$ ms..  

\begin{figure}[h!]
\centering
\includegraphics[scale=0.205]{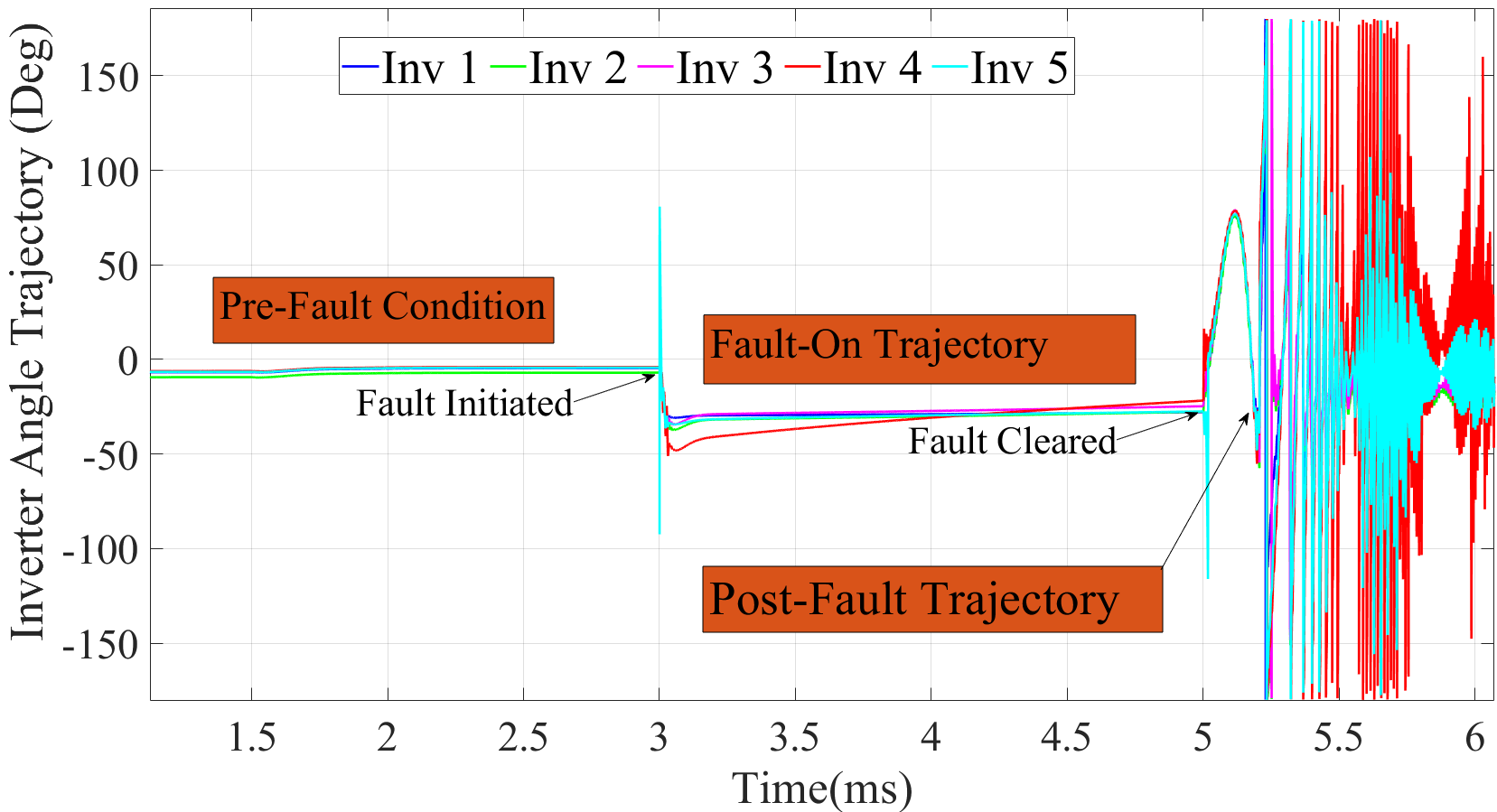}
\caption{Power Angle Trajectory of Parallel Inverters During Pre-fault, Fault-on and Post-fault conditions: Unstable Condition}
\label{fig:ut}
\end{figure}

In Fig. \ref{fig:ut}, it is seen that the non-uniform aggregated inverters lose synchronism if the fault is cleared after $5$ ms. It is also discovered that inverters with the largest apparent power fail synchronism faster than the others. Under this low-inertia condition, the electrical outage happens in the distribution feeder faster than the uniform operation of inverters. After running several cases, we estimate the CCT to be around $1.57$ ms. Also, we have found that the CCT for non-uniform inverters is smaller by $0.15$ ms after comparing with uniform inverters. The manufacture inverter fault duration based on critical clearing time (CCT) ranges from 1.1 to 4.25 ms \cite{keller2010understanding}. We have also found the CCT around this range 1.57ms-3.87ms depending on different line parameters, apartment power limit of multi-inverters, and connected Thevenin equivalent source.

\section{Conclusion}

In this work, the angular stability for parallel-connected grid following inverter has been assessed. We demonstrated that line fault in the neighbor feeder causes the distributed parallel inverters to lose synchronism and experience angular instability in a low inertia weak grid. The results showed that the voltage angle trajectory depends on the inverter's apparent power and line impedances when no voltage support exists in PCC due to the weak grid condition or inherent operation of current source inverters. Also, the current injection increases and plateaus at a particular limit, which depends upon the dc side inductor, inverter apparent power, and $X/R$ ratio. If uniformity is maintained for parallel-connected inverters in terms of power ratings and line impedances, the inverter loses synchronism in a narrow-angle gap. However, if uniformity is not maintained, inverter angular instability is caused by a wide margin of angle gap. In such a case, high-rated inverter injection current limits reach faster than others and shut off to prevent the unsafe operation of power-electronics devices. Under this condition, the $V_{PCC}$ reduces further, and the current injections of connected inverters increase further. Eventually, other inverters reach the unstable point faster than the inverters under uniform conditions. Finally, the critical clearing time is significantly compromised under this condition.

%\section*{Acknowledgement}
%This material is based upon work supported by the National Science Foundation, USA under Grant ECCS-2033927.

\bibliographystyle{ieeetr}
\bibliography{reference}

\end{document}